\documentclass[twoside,preprintnumbers,amsmath,amssymb,pacs,twocolumn,nopacs,nofootinbib]{revtex4}
\usepackage{amsmath,amssymb,url}
\usepackage{graphicx,pslatex,subfigure}

\usepackage{braket,euscript}
\usepackage{fancyvrb}

\newcommand{\p}{\partial}

\begin{document}

\title{{\bf An exact nilpotent non-perturbative BRST symmetry for the Gribov-Zwanziger action in the linear covariant gauge}}
\author{M.~A.~L.~Capri}
\email{caprimarcio@gmail.com}
\affiliation{Departamento de F\'{\i}sica Te\'{o}rica,
Instituto de F\'{\i}sica, UERJ - Universidade do Estado do Rio de Janeiro, Rua S\~ao Francisco Xavier 524, 20550-013, Maracan\~a, Rio de Janeiro, Brasil}\author{D.~Dudal}
\email{david.dudal@kuleuven-kulak.be}
\affiliation{KU Leuven Campus Kortrijk - KULAK, Department of Physics, Etienne Sabbelaan 53, 8500 Kortrijk, Belgium}
\affiliation{Ghent University, Department of Physics and Astronomy, Krijgslaan 281-S9, 9000 Gent, Belgium}
\author{D.~Fiorentini}
\email{diegofiorentinia@gmail.com}
\affiliation{Departamento de F\'{\i}sica Te\'{o}rica,
Instituto de F\'{\i}sica, UERJ - Universidade do Estado do Rio de Janeiro, Rua S\~ao Francisco Xavier 524, 20550-013, Maracan\~a, Rio de Janeiro, Brasil}\author{M.~S.~Guimaraes}
\email{msguimaraes@uerj.br}
\affiliation{Departamento de F\'{\i}sica Te\'{o}rica,
Instituto de F\'{\i}sica, UERJ - Universidade do Estado do Rio de Janeiro, Rua S\~ao Francisco Xavier 524, 20550-013, Maracan\~a, Rio de Janeiro, Brasil}\author{I.~F.~Justo}
\email{igorfjusto@gmail.com}
\affiliation{Departamento de F\'{\i}sica Te\'{o}rica,
Instituto de F\'{\i}sica, UERJ - Universidade do Estado do Rio de Janeiro, Rua S\~ao Francisco Xavier 524, 20550-013, Maracan\~a, Rio de Janeiro, Brasil}\affiliation{Ghent University, Department of Physics and Astronomy, Krijgslaan 281-S9, 9000 Gent, Belgium}
\author{A.~D.~Pereira}
\email{aduarte@if.uff.br}
\affiliation{UFF $-$ Universidade Federal Fluminense, Instituto de F\'{\i}sica, Campus da Praia Vermelha, Avenida General Milton Tavares de Souza s/n, 24210-346, Niter\'oi, RJ, Brasil.}
\author{B.~W.~Mintz}
\email{bruno.mintz.uerj@gmail.com}
\affiliation{Departamento de F\'{\i}sica Te\'{o}rica,
Instituto de F\'{\i}sica, UERJ - Universidade do Estado do Rio de Janeiro, Rua S\~ao Francisco Xavier 524, 20550-013, Maracan\~a, Rio de Janeiro, Brasil}\author{L.~F.~Palhares}
\email{leticiapalhares@gmail.com}
\affiliation{Departamento de F\'{\i}sica Te\'{o}rica,
Instituto de F\'{\i}sica, UERJ - Universidade do Estado do Rio de Janeiro, Rua S\~ao Francisco Xavier 524, 20550-013, Maracan\~a, Rio de Janeiro, Brasil}\author{R.~F.~Sobreiro}
\email{sobreiro@if.uff.br}
\affiliation{UFF $-$ Universidade Federal Fluminense, Instituto de F\'{\i}sica, Campus da Praia Vermelha, Avenida General Milton Tavares de Souza s/n, 24210-346, Niter\'oi, RJ, Brasil.}
\author{S.~P.~Sorella}
\email{silvio.sorella@gmail.com}
\affiliation{Departamento de F\'{\i}sica Te\'{o}rica,
Instituto de F\'{\i}sica, UERJ - Universidade do Estado do Rio de Janeiro, Rua S\~ao Francisco Xavier 524, 20550-013, Maracan\~a, Rio de Janeiro, Brasil}

\begin{abstract}
We point out the existence of a non-perturbative exact nilpotent BRST symmetry for the Gribov-Zwanziger action in the Landau gauge. We then put forward a manifestly BRST invariant resolution of the Gribov gauge fixing ambiguity in the linear covariant gauge.
\end{abstract}
\maketitle

\section{Introduction}
The Gribov-Zwanziger framework  \cite{Gribov:1977wm,Zwanziger:1989mf} is a non-perturbative approach to face the hard problem of understanding the behavior of Yang-Mills theories in the infrared region, where standard perturbation theory cannot be applied. It takes into account the existence of Gribov copies\footnote{For pedagogical reviews of the Gribov problem, see \cite{Sobreiro:2005ec,Vandersickel:2012tz}. }  \cite{Gribov:1977wm}, resulting in a modification of the Faddeev-Popov quantization formula for the Euclidean functional integral. Gribov copies are present whenever the gauge fixing condition allows multiple solutions, a very generic feature as shown by \cite{Singer:1978dk}. So far, a non-trivial set of results has been obtained from this approach, ranging from the gluon and ghost two-point functions \cite{Dudal:2007cw,Dudal:2008sp,Dudal:2011gd}, to the glueball spectrum \cite{Dudal:2010cd,Dudal:2013wja}, to thermodynamic quantities and phase transitions \cite{Canfora:2015yia,Canfora:2013kma,Canfora:2013zna,Lichtenegger:2008mh,Fukushima:2012qa,Fukushima:2013xsa,Su:2014rma}, to supersymmetric theories  \cite{Capri:2014tta,Capri:2014xea} and to the case where Higgs matter fields are present \cite{Capri:2012ah}. Nevertheless, the important  issue of the BRST symmetry still lacks a simple answer, see  \cite{Dudal:2009xh,Sorella:2009vt,Baulieu:2008fy,Capri:2010hb,Dudal:2012sb,Dudal:2014rxa,Capri:2012wx,Pereira:2013aza,Pereira:2014apa,Capri:2014bsa,Tissier:2010ts,Serreau:2012cg,Serreau:2015yna,Lavrov:2013boa,Moshin:2015gsa,Schaden:2014bea,Cucchieri:2014via} for an overview of the on-going discussion. In the present paper we propose a manifestly BRST invariant formulation of the Gribov-Zwanziger framework, resulting in the existence of a non-perturbative exact  BRST symmetry.   We limit ourselves here to outline the main steps of our reasoning, postponing all details to a longer and complete work.

\section{The original Gribov-Zwanziger action in the Landau gauge}
The framework \cite{Gribov:1977wm,Zwanziger:1989mf}, applied to $SU(N)$ gauge theories in Euclidean space-time,  implements the restriction  of the path integral to the Gribov region $\Omega$ in the Landau gauge, $\partial_\mu A^a_\mu=0$, namely
\begin{equation}
\Omega = \{ \; A^a_\mu|  \;  \partial_\mu A^a_\mu =0,  \;    {\mathcal M}^{ab}(A) > 0 \; \}    \;, \label{om}
\end{equation}
where ${\mathcal{M}}^{ab}$ is the Faddeev-Popov operator
\begin{equation}
\mathcal{M}^{ab}=-\delta^{ab}\partial^2+gf^{abc}A^{c}_{\mu}\partial_{\mu},\,\,\,\, \mathrm{with}\,\,\,\, \partial_{\mu}A^{a}_{\mu}=0\,.
\label{intro0}
\end{equation}
According to \cite{Gribov:1977wm,Zwanziger:1989mf},  for the partition function of quantized Yang-Mills theory we write
\begin{equation}
\EuScript{Z}=\int_{\Omega} \left[\EuScript{D}\mathbf{A}\right]\; \delta(\partial A^a) \; \det({\cal M})\;\mathrm{e}^{- S_{\mathrm{YM}} }\,.
\label{intro1}
\end{equation}
The restriction of the domain of integration to the region $\Omega$ can be effectively implemented by adding to the starting action an additional non-local term $H(A)$, known as the horizon function. More precisely \cite{Gribov:1977wm,Zwanziger:1989mf}
\begin{eqnarray}
&&\int_{\Omega} \left[\EuScript{D}\mathbf{A}\right]\; \delta(\partial A^a) \; \det({\cal M})\;\mathrm{e}^{- S_{\mathrm{YM}} }\nonumber\\&&  = \int \left[\EuScript{D}\mathbf{A}\right]\; \delta(\partial A^a) \; \det({\cal M})\;\mathrm{e}^{- \left( S_{\mathrm{YM}} +\gamma^4H(A) - 4V\gamma^4(N^2-1) \right) }
\label{gz1}
\end{eqnarray}
where
\begin{equation}
H(A)=g^2\int d^4xd^4y~f^{abc}A^{b}_{\mu}(x)\left[\mathcal{M}^{-1}(x,y)\right]^{ad}f^{dec}A^{e}_{\mu}(y)\,,
\label{intro3}
\end{equation}
with $\left[\mathcal{M}^{-1}\right]$ denoting the inverse of the Faddeev-Popov operator, see eq.~\eqref{intro0}.
The mass parameter  $\gamma^2$ appearing in expression \eqref{gz1}  is known as the Gribov parameter. It  is determined in a self-consistent way by the  gap equation \cite{Zwanziger:1989mf}
\begin{equation}
\langle H \rangle = 4V(N^2-1)\,,
\label{intro4}
\end{equation}
where the vacuum expectation value $\langle H \rangle$ has to be evaluated with the measure defined in eq.~\eqref{gz1}; $V$ denotes the space-time volume. Expression \eqref{gz1} can be cast in a more suitable form by introducing a set of commuting $({\bar \phi}, \phi)$ and anticommuting $(\omega, {\bar \omega})$ auxiliary fields  \cite{Zwanziger:1989mf}, namely
\begin{equation}
\int_{\Omega} \left[\EuScript{D}\mathbf{A}\right]\; \delta(\partial A^a) \; \det({\cal M})\;\mathrm{e}^{- S_{\mathrm{YM}} } = \int  \left[\EuScript{D}{{\Phi}}\right]\;\mathrm{e}^{- \left( S_{GZ}  - 4V\gamma^4(N^2-1) \right)},
\label{gzact}
\end{equation}
where $\Phi$ refers to all fields present and $S_{GZ}$ stands for the Gribov-Zwanziger action\footnote{We employ here a short-hand notation, namely
$
{\bar \phi} {{\cal M}}(A) \phi  =   {\bar \phi}^{ac}_\mu  {{\cal M}}(A)^{ab} \phi^{bc}_\mu \;,
{\bar \omega} {{\cal M}}(A) \omega =   {\bar \omega}^{ac}_\mu  {{\cal M}}(A)^{ab} \omega^{bc}_\mu \;,
\gamma^2 A ({\bar \phi} + \phi) =   g \gamma^2 f^{abc}A^a_\mu ({\bar \phi}^{bc}_\mu + \phi^{bc}_\mu)  \;. \label{not}
$
}
\begin{equation}
S_{GZ} = S_{FP} + \int d^4x\;  \left( {\bar \phi} {{\cal M}}(A) \phi - {\bar \omega} {{\cal M}}(A) \omega + \gamma^2 A ({\bar \phi} + \phi)   \right)\;, \label{gzact}
\end{equation}
with $S_{FP}$ being the Faddeev-Popov action in the Landau gauge
\begin{equation}
S_{FP} = S_{YM} + \int d^4x \left( b^a \partial_\mu A^a_\mu + {\bar c}^a \partial_\mu D^{ab}_\mu c^b \right)    \;.   \label{fp}
\end{equation}
Notice that the gap equation \eqref{intro4} can be rewritten as
\begin{equation}
\frac{\partial{\mathcal E}_v}{\partial \gamma^2}=0 \;, \qquad  e^{-V{\mathcal E}_v} = \int  \left[\EuScript{D}{{\Phi}}\right]\;\mathrm{e}^{- \left( S_{GZ}  - 4V\gamma^4(N^2-1) \right)}  \;, \label{gapr}
\end{equation}
where ${\mathcal E}_v$ denotes the vacuum energy.  As already mentioned, till now, a simple resolution of the issue of the BRST symmetry for the action \eqref{gzact} is still lacking.

\noindent One important property which should be underlined here is that, as observed in \cite{Dudal:2008sp}, the Gribov region $\Omega$ does not support anymore infinitesimal gauge transformations. If one performs an infinitesimal gauge transformation of a generic field $A_\mu$ belonging to $\Omega$, the resulting transformed field lies outside the region $\Omega$. From this simple argument, one easily understands that the restriction of the functional integral to the region $\Omega$ might give rise to  possible incompatibilities with the standard BRST symmetry.

\section{Warming up: a non-perturbative exact BRST symmetry for the Gribov-Zwanziger action in the Landau gauge}
The previous observation has led us to consider a non-local gauge invariant transverse field $A_{\mu}^h$, $\partial_\mu A^h_\mu=0$, obtained by minimizing the auxiliary functional $\mathrm{Tr}\int d^{4}x\,A_{\mu }A_{\mu
}$  along the gauge orbit of $A_\mu$, cf.~\cite{Zwanziger:1990tn, Lavelle:1995ty,Capri:2005dy} and Appendix \ref{apb},
\begin{eqnarray}
A_{\mu }^{h} &=&P_{\mu\nu} \left( A_{\nu }-ig\left[ \frac{\partial A}{\partial
^{2}},A_{\nu }\right] +\frac{ig}{2}\left[ \frac{\partial A}{\partial ^{2}}%
,\partial _{\nu }\frac{\partial A}{\partial ^{2}}\right] \right)
+O(A^{3})  \nonumber \\
&=&A_{\mu }-\frac{\partial _{\mu }}{\partial ^{2}}\partial A+ig\left[ A_{\mu
},\frac{1}{\partial ^{2}}\partial A\right] +\frac{ig}{2}\left[ \frac{1}{%
\partial ^{2}}\partial A,\partial _{\mu }\frac{1}{\partial ^{2}}\partial
A\right]\nonumber\\&& +ig\frac{\partial _{\mu }}{\partial ^{2}}\left[ \frac{\partial
_{\nu }}{\partial ^{2}}\partial A,A_{\nu }\right]   +i\frac{g}{2}\frac{\partial _{\mu }}{\partial ^{2}}\left[ \frac{\partial A%
}{\partial ^{2}},\partial A\right] +O(A^{3})\,,  \label{hhh3g}
\end{eqnarray}
with $P_{\mu\nu}=\left(\delta _{\mu \nu }-\frac{\partial _{\mu }\partial
_{\nu }}{\partial ^{2}}\right) $ the transverse projector.

\noindent Expression \eqref{hhh3g} is left invariant by infinitesimal gauge transformations order by order. Moreover, looking at eq.~\eqref{hhh3g}, one realizes that a divergence $\partial A$ is present in all higher order terms. As a consequence, we can rewrite Zwanziger's horizon function $H(A)$ in terms of the invariant field $A^h$ as
\begin{equation}
H(A) = H(A^h) - R(A) (\partial A)
\end{equation}
where $R(A)(\partial A)$ is a short-hand notation, $R(A) (\partial A)= \int d^4x d^4y R^a(x,y) (\partial A^a)_y$, $R(A)$ being an infinite non-local power series of $A_\mu$. Therefore, for the Gribov-Zwanziger action, we may write, omitting color indices for brevity,
\begin{eqnarray}
S_{GZ} &=& S_{YM} + \int d^4x \left( b \partial_\mu A_\mu + {\bar c} \partial_\mu D_\mu c \right)  + \gamma^4 H(A)     \nonumber \\
&&\hspace{-1cm} = ~S_{YM} + \int d^4x \left( b \partial_\mu A_\mu + {\bar c} \partial_\mu D_\mu c  \right)  + \gamma^4 H(A^h) -\gamma^4 R(A) (\partial A)     \nonumber \\
& = & S_{YM} + \int d^4x \left( b^{h} \partial_\mu A_\mu + {\bar c} \partial_\mu D_\mu c \right)  + \gamma^4 H(A^h)      \;,
\label{gzh2}
\end{eqnarray}
where the new field $b^h$ stands for
\begin{equation}
b^h = b - \gamma^4 R(A)   \;. \label{bh}
\end{equation}
The use of the field $b^h$ enables us to write down an exact nilpotent non-perturbative BRST transformation. Rewriting the Gribov-Zwanziger action by using the auxiliary  fields $({\bar \phi}, \phi, \omega, {\bar \omega})$, i.e.
\begin{eqnarray}
S_{GZ} &=& S_{YM} + \int d^4x \left( b^h \partial_\mu A_\mu + {\bar c} \partial_\mu D_\mu c \right) \nonumber\\&& \hspace{-0.9cm}+ \int d^4x\;  \left( {\bar \phi} {{\cal M}}(A^h) \phi - {\bar \omega} {{\cal M}}(A^h) \omega + \gamma^2 A^h ({\bar \phi} + \phi)   \right), \label{gzh3}
\end{eqnarray}
it becomes clear that expression \eqref{gzh3} is left invariant by the nilpotent non-perturbative BRST transformation
\begin{equation}
s_{\gamma^2} =  s + \delta_{\gamma^2}\,,\qquad  s_{\gamma^2}^2=0\;, \qquad
s_{\gamma^2} S_{GZ} = 0 \;. \label{gzinv}
\end{equation}
In eqs.~\eqref{gzinv}, the operator $s$ stands for  the usual BRST operator
\begin{eqnarray}
s A^a_\mu &=&  - D_\mu^{ab} c^b \;,~ s c^a = \frac{g}{2} f^{abc} c^b c^c   \;,~ s {\bar c}^a  =  b^a  \;, ~s b^a = 0      \;, \nonumber \\
s \phi_\mu^{ab} & = & \omega_\mu^{ab} \;, ~ s \omega_\mu^{ab}=0 \;, ~ s {\bar \omega}_\mu^{ab} =   {\bar \phi}_\mu^{ab}   \;, ~ s{\bar \phi}_\mu^{ab} =0   \;, \label{brst3}
\end{eqnarray}
while
\begin{eqnarray}
\delta_{\gamma^2} {\bar c}^a & = & - \gamma^4 R^a(A)  \;,\quad \delta_{\gamma^2}b^a = \gamma^4 sR^a(A) \;,\nonumber\\
\delta_{\gamma^2} {\bar \omega}_\mu^{ac}& =&  \gamma^2 gf^{kbc} A_\mu^{h,k}\left[{\cal M}^{-1}(A^h)\right]^{ba}\;,~\delta_{\gamma^2}(\text{rest})=0\;.
\end{eqnarray}
The operators $(s,\delta_{\gamma^2})$ obey the nice algebra
\begin{equation}\label{alg}
  \{s,\delta_{\gamma^2}\}= s^2=\delta_{\gamma^2}^2=s_{\gamma^2}^2=0
\end{equation}
and clearly, for $\gamma^2\to0$ we have $s_{\gamma^2}\to s$.

\noindent The  operator $s_{\gamma^2}$  is a genuine  non-perturbative BRST operator, as it depends explicitly on the non-perturbative Gribov parameter $\gamma^2$.

\noindent Thanks to $s_{\gamma^2}$, we can write down non-perturbative Ward identities which clarify the origin of the breaking of the  standard BRST operator. From the non-perturbative exact Slavnov-Taylor Ward identities
\begin{equation}
\langle s_{\gamma^2} \left( {\bar c} \Lambda \right) \rangle  = 0 \;, \label{np}
\end{equation}
where $\Lambda$ has ghost number zero, it follows that the operator $s$ will always acquire a breaking term proportional to $\gamma^2$, namely
\begin{equation}
\langle s \left( {\bar c} \Lambda \right) \rangle  =  - \langle \delta_{\gamma^2} \left( {\bar c} \Lambda \right) \rangle \;. \label{npt}
\end{equation}
This equation gives a clear and simple understanding of the origin of the breaking of the standard BRST symmetry $s$. It states that $s$ is always plagued by breaking terms which are proportional to the non-perturbative Gribov parameter and it signals that, in presence of the Gribov horizon, the BRST operator $s$ has to be replaced by the non-perturbative one $s_{\gamma^2}$. It is the breaking of $s$ that has also been signalled recently on the lattice \cite{Cucchieri:2014via}. We will come back to this in a more detailed forthcoming paper.

\noindent Moreover, we notice that
\begin{equation}\label{nn1}
  \frac{\p S_{GZ}}{\p \gamma^2}\neq s_{\gamma^2}(\text{something})\,,
\end{equation}
indicating that the Gribov parameter $\gamma^2$ is \emph{not} akin to a gauge parameter. As such, it will enter physical quantities. With physical quantities, we mean the colorless gauge invariant operators which are immediately seen to belong to the cohomology of the new BRST operator $s_{\gamma^2}$.

\section{Gribov problem in the linear covariant gauge and its BRST invariant resolution}
Having found a non-perturbative exact nilpotent symmetry of the Gribov-Zwanziger action in the Landau gauge, we move to the linear covariant gauges. We shall proceed by staying as close as possible to the BRST construction of the gauge-fixing, i.e.~by
defining it as an exact non-perturbative variation, by employing the nilpotent operator $s_{\gamma^2}$ introduced before.
Moreover, this construction will be linked to the introduction of a suitable region $\Omega^h$ in field space which shares many properties of the Gribov region $\Omega$ of the Landau gauge.

\noindent Thus, according to the general BRST procedure for the gauge-fixing, we   write down the following $s_{\gamma^2}$-invariant action
\begin{equation}
S_{GZ}^{LCG} = S_{FP}^{h} + \int d^4x\;  \left( {\bar \phi} {{\cal M}}(A^h) \phi - {\bar \omega} {{\cal M}}(A^h) \omega + \gamma^2 A^h ({\bar \phi} + \phi)   \right)\;, \label{fp88a}
\end{equation}
with
\begin{eqnarray}
S_{FP}^{h} &=& S_{YM} + s_{\gamma^2}\int d^4x \left( \bar c \p_\mu A_\mu - \frac{\alpha}{2} \bar cb^h\right)\nonumber\\       \label{fpLCG}
&=& S_{YM}+ \int d^4x \left( b^h \p_\mu A_\mu - \frac{\alpha}{2}b^h b^h +\bar c \p_\mu D_\mu c\right)
\end{eqnarray}
Expression \eqref{fp88a}  naturally generalizes the Gribov-Zwanziger action of the Landau gauge to an arbitrary linear covariant gauge in a manifestly non-perturbative BRST invariant way, namely
\begin{equation}
s_{\gamma^2} S_{GZ}^{LCG}=0   \;. \label{sinvlcg}
\end{equation}
The action \eqref{fp88a} reduces precisely to the Gribov-Zwanziger action in the limit $\alpha\rightarrow 0$
 \begin{equation}
 S_{GZ}^{LCG} \big|_{\alpha=0} = S_{GZ}   \;,
 \end{equation}
while yielding  the usual  action of the linear covariant gauge when $\gamma^2=0$, i.e.
\begin{equation}
S_{GZ}^{LCG} \big|_{\gamma^2=0} = S_{FP} = S_{YM}+ \int d^4x \left( b \p_\mu A_\mu - \frac{\alpha}{2}b b +\bar c \p_\mu D_\mu c\right)   \;, \label{lcg11}
\end{equation}
Expression \eqref{lcg11} is nothing but the Faddeev-Popov action of the linear covariant gauges
\begin{equation}\label{fp2}
  \p_\mu A_\mu= \alpha b \;,
\end{equation}
where $\alpha$ stands for the gauge parameter and $b$ for the Lagrange multiplier.

\noindent Since in expression \eqref{fp88a} the gauge parameter $\alpha$ is coupled to a $s_{\gamma^2}$-exact quantity, expectation values of $s_{\gamma^2}$-invariant quantities will not depend on $\alpha$. In particular, this  will be the case for the  dynamical mass scale $\gamma^2$. As we shall see at the end of this section, the independence of  $\gamma^2$ from $\alpha$ is a consequence of the fact  that $\gamma^2$ is now determined  by the gauge invariant horizon condition
\begin{eqnarray}
\frac{\partial{\mathcal E}_v}{\partial \gamma^2}& = & 0  \Rightarrow  \langle H(A^h) \rangle = 4V(N^2-1)\, \nonumber \\   \mathrm{e}^{-V{\mathcal E}_v} & = &  \int  \left[\EuScript{D}{{\Phi}}\right]\;\mathrm{e}^{- \left( S_{GZ}^{LCG}  - 4V\gamma^4(N^2-1) \right)}  \;, \label{gaprlcg}
\end{eqnarray}
where use has been made of the identity
\begin{equation}
 \int  \left[\EuScript{D}{{\Phi}}\right]\;  \frac{\delta }{\delta b} \left( {\mathcal F(A)}  \mathrm{e}^{- \left( S_{GZ}^{LCG}  - 4V\gamma^4(N^2-1) \right)}  \right) =0 \;,
\end{equation}
valid for an arbitrary quantity ${\mathcal F(A)}$.

\noindent It is also interesting to note that, integrating out the field $b^h$ in expression \eqref{fp88a}, one gets the nice equation
\begin{equation}
\int d^4x \left( b^h \p_\mu A_\mu - \frac{\alpha}{2}b^h b^h \right)  \Rightarrow \int d^4x   \frac{1}{2\alpha} (\p_\mu A_\mu)^2  \;.
\end{equation}
We point out that, recently, the linear covariant gauges have been studied in lattice numerical simulations by \cite{Cucchieri:2009kk,Bicudo:2015rma} or with functional methods by \cite{Aguilar:2007nf,Siringo:2014lva,Aguilar:2015nqa,Huber:2015ria}. It is worth underlining  that the tree level gluon propagator \cite{Capri:2015pja} stemming from expression \eqref{fp88a}  turns out to be in qualitative agreement with the available lattice numerical simulations \cite{Cucchieri:2009kk,Bicudo:2015rma}, exhibiting an infrared suppression in the gluon sector. A more detailed analysis will involve taking into account additional $d=2$ condensates, following \cite{Dudal:2008sp}. Let us provide a geometrical understanding of the action \eqref{fp88a} by showing that it enables one to eliminate infinitesimal gauge copies.

\noindent The Faddeev-Popov operator for general $\alpha$ reads
\begin{eqnarray}\label{fp3}
 \mathcal{M}^{ab}(A)=-\p_\mu D_\mu^{ab}=-\p_\mu(\delta^{ab}\p_\mu-gf^{abc}A_\mu^c)\nonumber\\=-\delta^{ab}\p^2+\alpha gf^{abc}b^{c}+gf^{abc}A_\mu^c\p_\mu\,.
\end{eqnarray}
Infinitesimal Gribov copies will appear whenever
\begin{equation}\label{fp4}
  \mathcal{M}^{ab}(A)\zeta^b=0\,,
\end{equation}
with $\zeta^a$ a normalizable zero mode, in which case $A_\mu^a- D_\mu^{ab}\zeta^b$ also fulfills condition \eqref{fp2} if $A_\mu^a$ does.

\noindent Unlike the case of the Landau gauge, we notice that, when $\alpha\neq0$,  the partial derivative $\p$ and the covariant one $D$ do not commute.  As a consequence, the Faddeev-Popov operator in eq.~\eqref{fp3} is not Hermitian. The Hermiticity of $\mathcal{M}^{ab}$ plays an important role in the original Gribov-Zwanziger analysis. Let us therefore consider
\begin{equation}\label{fp5}
  \mathcal{M}^{ab}(A^h)=-\p_\mu(\delta^{ab}\p_\mu-gf^{abc}A_\mu^{h,c})\,,
\end{equation}
with $A^h$ the gauge invariant field defined in eq.~\eqref{hhh3g}. By construction, the operator $\mathcal{M}(A^h)$ in eq.~\eqref{fp5} is gauge invariant order by order and Hermitian, thanks to the transversality of $A^h$.  It thus makes sense to define the region
\begin{equation}
\Omega^h = \{ \; A_\mu|   \p_\mu A_\mu^a=\alpha b^a,\; \partial_\mu A^h_\mu =0,  \;    {\mathcal M}^{ab}(A^h) > 0 \; \}    \;. \label{fp7}
\end{equation}
The region $\Omega^h$  shares the important properties of the Gribov region $\Omega$ of the Landau gauge of being convex and bounded in all directions  \cite{Dell'Antonio:1991xt}. Those properties follow from the linearity of the operator ${\mathcal M}^{ab}(A^h)$ in the field $A^h$.

\noindent Let us recall that the Landau gauge is, as far as we know, the only gauge for which it has been proven that every gauge orbit crosses at least once the Gribov region $\Omega$
\cite{Dell'Antonio:1991xt,semenov}, i.e.~a gauge field configuration located outside of the region $\Omega$ is a copy of some configuration located within $\Omega$. The essential ingredient in the proof of \cite{Dell'Antonio:1991xt,semenov} is that the functional $\mathrm{Tr}\int d^{4}x\,A_{\mu }A_{\mu}$  achieves its absolute minimum along the gauge orbit of $A$, and this for an arbitrary starting gauge configuration $A$. Said otherwise, the search for the minima  along the gauge orbit can be regarded as a pure mathematical problem for the functional $\mathrm{Tr}\int d^{4}x\,A_{\mu }A_{\mu}$, not related to the particular gauge condition obeyed by the configuration $A$.  Actually, it turns out that the functional $\mathrm{Tr}\int d^{4}x\,A_{\mu }A_{\mu}$ has many relative minima along the gauge orbit before attaining its absolute minimum. The set of the relative minima of $\mathrm{Tr}\int d^{4}x\,A_{\mu }A_{\mu}$ is precisely the Gribov region $\Omega$. The proof of \cite{Dell'Antonio:1991xt,semenov} shows thus that, given an arbitrary gauge configuration $A$, it is always possible to introduce a related transverse field $A^h$ through the process of  minimization of the functional $\mathrm{Tr}\int d^{4}x\,A_{\mu }A_{\mu}$ along the gauge orbit of $A$. Any configuration $A^h$  can be identified with a local minimum of the functional $\mathrm{Tr}\int d^{4}x\,A_{\mu }A_{\mu}$, while any such minimum is left invariant by infinitesimal gauge transformations. Our construction of a non-perturbative BRST operator is possible with any $A^h$, but for our purposes we use the unique order by order representation given in eq.~\eqref{hhh3g}. These considerations make the region $\Omega^h$ a suitable candidate to integrate over.

\noindent Let us proceed by showing that the use of the region \eqref{fp7} enables us to  eliminate a large class of infinitesimal gauge copies from the partition function. This proposition borrows from an earlier insight of some of us in \cite{Sobreiro:2005vn,Capri:2015pja}, where only the transverse component $A^T_\mu$, $A^T_\mu = (\delta_{\mu\nu} - \frac{\partial_\mu \partial_\nu}{\partial^2} )A_\nu$,  was considered instead of the complete invariant gauge field $A^h$.

\noindent Following \cite{Sobreiro:2005vn,Capri:2015pja}, let us assume that $\zeta^a$ is a zero mode of the Faddeev-Popov operator \eqref{fp3} having a Taylor expansion in $\alpha$,
\begin{equation}
\zeta^a = \sum_{n=0}^{\infty} \alpha^n\zeta_n^a.
\end{equation}
Let us decompose the gauge field $A^a_\mu$ according to
\begin{equation}
A_\mu =  A^h_\mu +\tau_\mu \;, \qquad \partial_\mu \tau_\mu = \alpha b \;, \label{t1}
\end{equation}
so that, in view of eq.~\eqref{t1}, we can write
\begin{equation}
\tau_\mu= \sum_{n=0}^{\infty} \alpha^{n+1}\tau_{\mu}^{n}  = \alpha {\hat \tau}_\mu \;, \label{t2}
\end{equation}
since $\tau_\mu$ has to vanish in the limit $\alpha \rightarrow 0$.
If $A_\mu \in \Omega^h$, we can write
\begin{eqnarray}
\zeta^a &=&- g \left[\mathcal{M}(A^h)^{-1}\right]^{ad}f^{dbc}\p_\mu\left(\tau_\mu^b \zeta^c\right)\nonumber\\&=&- g\alpha\left[\mathcal{M}(A^h)^{-1}\right]^{ad}f^{dbc}\p_\mu\left({\hat \tau}_\mu^b\zeta^c\right),
\end{eqnarray}
or, expanding in powers of $\alpha$,
\begin{eqnarray}
\sum_n \alpha^n \zeta_n^a= -\sum_{n} g\alpha^{n+1}\left[\mathcal{M}(A^h)^{-1}\right]^{ad}f^{dbc}\p_\mu\left(\zeta_n ^c{\hat \tau}_\mu^{b}\right)
\end{eqnarray}
Matching orders of $\alpha$ shows that the $n^{\text{th}}$ order coefficient $\zeta_n^a$ is proportional to the $(n-1)^{\text{th}}$. Since for the first coefficient we find $\zeta_0^a=0$, we immediately find $\zeta_n^a=0$, and thus $\zeta^a=0$. Said otherwise, all zero modes that possess a Taylor expansion around  $\alpha=0$, are automatically vanishing. As such, the restriction to $\Omega^h$ excludes at least the set of infinitesimally connected gauge copies related to the aforementioned zero modes.

\noindent We proceed by implementing $\mathcal{M}^h\equiv\mathcal{M}^{ab}(A^h)>0$ into the path integral. We rely on the so-called Gribov no-pole condition \cite{Gribov:1977wm}, whose all order implementation can be found in \cite{Capri:2012wx}. For any external field $A^h$, we can use Wick's theorem to invert the operator $\mathcal{M}^{ab}(A^h)$ in any dimension $d$.  Denoting by $\mathcal{G}^{ab}(A^h,p^2)= \langle p | \frac{1}{\mathcal{M}^{ab}(A^h)} | p \rangle $ the Fourier-transform of the inverse of $\mathcal{M}^{ab}(A^h)$, one introduces the so-called Gribov form factor \cite{Capri:2012wx} $\sigma(A^h,p^2)$ through
\begin{equation}\label{nop0}
  \mathcal{G}^{ab}(A^h, p^2)=\frac{\delta^{ab}}{N^2-1}\mathcal{G}^{cc}(A^h,p^2)=\frac{\delta^{ab}}{N^2-1}\frac{1+\sigma(A^h,p^2)}{p^2}.
\end{equation}
Repeating the procedure outlined in \cite{Capri:2012wx}, it follows that at zero momentum
\begin{eqnarray}\label{nop}
\sigma(A^h,0)&=&\\&&\hspace{-2cm}-\frac{g^2}{Vd(N^2-1)}\int \frac{d^d k}{(2\pi)^d}\frac{d^d q}{(2\pi)^d} A_\mu^{h,ab}(-k) \left[(\mathcal{M}^{h})^{-1}\right]^{bc}_{k-q}A_\mu^{h,ca}(q).\nonumber
\end{eqnarray}
Comparison of eqns.~\eqref{intro3} and \eqref{nop} learns that $\sigma(A^h,0)=\frac{H(A^h)}{Vd(N^2-1)}$. We will concentrate on the zero momentum limit, since it is expected on general grounds\footnote{We can consider $\mathcal{M}^{ab}(A^h)$ as a perturbed system around $-\p^2$, which reaches its lowest eigenvalue at zero momentum. A few comments regarding this were made in \cite{Vandersickel:2012tz}. One can also check, a posteriori but explicitly, that the expectation value $\braket{\sigma(A^h,0)}$ is maximal.} that the smallest eigenvalue of $\mathcal{M}^{ab}(A^h)$ will carry no momentum, so it would be sufficient to avoid this eigenvalue becoming negative. At the level of expectation values, we can rewrite eq.~\eqref{nop} as
\begin{eqnarray}\label{nop2}
\mathcal{G}^h(p^2)= \braket{ \mathcal{G}^{aa}(A^h, p^2)}^{conn}=\frac{1}{p^2(1-\braket{\sigma(A^h,p^2)}^{1PI})},
\end{eqnarray}
so that we must impose at the level of the path integral $\braket{\sigma(A^h,0)}^{1PI}\leq 1$, or
\begin{eqnarray}\label{nop2b}
  \braket{H(A^h)}^{1PI} &\leq& Vd(N^2-1)\,.
\end{eqnarray}
We can add this constraint to the path integral measure with a step function. Via a saddle point evaluation in the thermodynamic limit \cite{Gribov:1977wm,Dudal:2014rxa}, one then finds
\begin{eqnarray}\label{last}
&&[\EuScript{D}{{\Phi}}] \theta[Vd(N^2-1)-H(A^h)]\mathrm{e}^{-S_{FP}^{h}}  \nonumber\\
&=& [\EuScript{D}{{\Phi}}]\int\frac{d\eta}{2\pi i\eta} \mathrm{e}^{-S_{FP}^{h}+\eta[Vd(N^2-1)-H(A^h)]}\nonumber\\
&\to&[\EuScript{D}{{\Phi}}]\mathrm{e}^{-S_{FP}^{h}+\eta^\ast[Vd(N^2-1)-H(A^h)]} \;,
\end{eqnarray}
where $S_{FP}^{h}$ stands for the expression given in eq.~\eqref{fpLCG}. The saddle point equation precisely amounts to eq.~\eqref{gaprlcg}, i.e.~the horizon condition with identification $\eta^\ast=\gamma^4$. As the horizon condition is writable in terms of the vacuum energy and since the only contributing diagrams to the latter are $1PI$ (see also \cite{Capri:2012wx}), it indeed follows that condition \eqref{nop2b} is met. As such, we do have excluded a large set of zero modes by effectively having imposed that $\mathcal{M}(A^h)> 0$ via the action \eqref{fp88a}. Upon introduction of the auxiliary fields $({\bar \phi}, \phi, \omega, {\bar \omega})$,  the latter is equivalent to  the action appearing in eq.~\eqref{fp88a}, given that eq.~\eqref{gaprlcg} holds.

\section{Conclusion}
For the first time, we have identified a non-perturbative nilpotent BRST symmetry for gauge theories quantized \`{a} la Gribov-Zwanziger, that is by further restricting the domain of integration in the path integral. This eliminates a large set of gauge copies and deeply affects the infrared low-momentum regime of the gauge theory. The new BRST operator $s_{\gamma^2}$ depends explicitly on the gauge invariant mass parameter $\gamma^2$ that is linked to the aforementioned restriction. As such, the operator $s_{\gamma^2}$ itself is intertwined with this geometric restriction.

The introduction of $s_{\gamma^2}$ opens up whole new strata of applications. We have already discussed a first one in this paper, namely a non-perturbative extension of the usual linear covariant gauge to a setting where the Gribov gauge fixing ambiguity is also faced in this gauge. Our setup generalizes to the Refined Gribov-Zwanziger approach \cite{Dudal:2008sp}, in which case we can make contact with the gauge invariant $d=2$ condensate $\braket{A^2_\text{min}}$, of important phenomenological interest \cite{Gubarev:2000eu,Gubarev:2000nz}. A renormalization analysis of the proposed framework is already in preparation, of relevance to explicit studies of propagators, spectrum and thermodynamics. Generalizations, compatible with the new non-perturbative BRST, to the matter sector are also possible. Moreover, it would also be  interesting to make contact with lattice studies of the linear covariant gauge, e.g.~to find out if a practical numerical implementation of our proposal exists. We are already studying a functional depending on the original gauge field $A_\mu$ and an auxiliary field $B_\mu$, with the property that the minimum occurs for $\p_\mu A_\mu=\alpha b$ (thus effectively implementing the linear covariant gauge) and for $B_\mu=A_\mu^h$ with $\mathcal{M}(B)\geq 0$. This could circumvent potential issues with the convergence of the series expression used in eq.~\eqref{hhh3g} to define $A^h$ in case of ``large'' gauge fields, while it would also open the road to simulation of our proposed non-perturbative linear covariant gauge. We will report on this in future work.

As a final but most crucial remark, we stress that no sacrifices have to be made w.r.t.~gauge invariance, even when the Gribov problem is taken into account. The physical content of the theory is described by the $s_{\gamma^2}$-cohomology, which can be studied along the lines of \cite{Piguet:1995er,Barnich:2000zw} upon localization of our approach, another matter of current investigation.

\section*{Acknowledgments}
The Conselho Nacional de Desenvolvimento Cient\'{i}fico e Tecnol\'{o}gico (CNPq-Brazil; R.~F.~S.~is a level PQ-2 researcher under the program Produtividade em Pesquisa , 308845/2012-9; M.~S.~G.~is a level PQ-2 researcher under the program Produtividade em Pesquisa, 307905/2014-4), the Coordena\c c\~ao de Aperfei\c coamento de Pessoal de N\'ivel Superior (CAPES) and the Pr\'o-Reitoria de Pesquisa, P\'os-Gradua\c c\~ao e Inova\c c\~ao (PROPPI-UFF) are acknowledged for financial support. L.~F.~P.~is supported by a BJT fellowship from the Brazilian program ``Ci\^encia sem Fronteiras'' (grant number 301111/2014-6).

\newpage
\appendix
\begin{widetext}
\section{A gauge invariant transversal gauge field} \label{apb}
As it will turn out, the construction of the transverse gauge field $A_\mu^h$ follows from  the minimization  of the functional
$f_{A}[u]$
\begin{equation}
f_{A}[u]\equiv \mathrm{Tr}\int d^{4}x\,A_{\mu }^{u}A_{\mu
}^{u}=\mathrm{Tr}\int d^{4}x\left( u^{\dagger }A_{\mu
}u+\frac{i}{g}u^{\dagger }\partial _{\mu }u\right) \left(
u^{\dagger }A_{\mu }u+\frac{i}{g}u^{\dagger }\partial _{\mu
}u\right)  \label{fa}
\end{equation}
along the gauge orbit of a given configuration $A_{\mu }$. To give a well defined mathematical meaning to expression \eqref{fa}, we shall require that both $A_{\mu }^{a}$ and the local gauge transformations, $u\in\mathcal{U}$, are square-integrable, i.e.~
\begin{equation}
||A||^{2}=\mathrm{Tr}\int d^{4}x\,A_{\mu }A{_{\mu }=}\frac{1}{2}\int
d^{4}xA_{\mu }^{a}A_{\mu }^{a}<+\infty \;,\qquad ||u^{\dagger }\partial {u}||^{2}=\mathrm{Tr}\int d^{4}x\,\left(
u^{\dagger }\partial _{\mu }u\right) \left( u^{\dagger }\partial
_{\mu }u\right) <+\infty \;. \label{norm1}
\end{equation}
Then, it has been shown \cite{Dell'Antonio:1991xt,semenov} that $f_A[u]$ reaches its absolute minimum along the gauge orbit
of $A_{\mu }$, i.e.~there exists a certain $h$ such that
\begin{eqnarray}
\delta f_{A}[h] &=&0\;,  \label{impl0} \\
\delta ^{2}f_{A}[h] &\ge &0\;,  \label{impl1} \\
f_{A}[h] &\le &f_{A}[u]\;,\;\;\;\;\;\;\;\forall \,u\in
\mathcal{U}\;. \label{impl2}
\end{eqnarray}
Following \cite{Zwanziger:1990tn,Lavelle:1995ty,Capri:2005dy}, we can work out the conditions (\ref{impl0}) and
(\ref{impl1}) in a series expansion. We set
\begin{equation}
v=he^{ig\omega }=he^{ig\omega ^{a}T^{a}}\;,  \label{set0}
\end{equation}
with
\begin{equation}
\left[ T^{a},T^{b}\right] =if^{abc}\;,\;\;\;\;\;\mathrm{Tr}\left( T^{a}T^{b}\right) =%
\frac{1}{2}\delta ^{ab}\;,  \label{st000}
\end{equation}
We first obtain\footnote{We refer to \cite{Capri:2005dy} for technical details.}
\begin{equation}
A_{\mu }^{v}=A_{\mu }^{h}+ig[A_{\mu }^{h},\omega ]+\frac{g^{2}}{2}[[\omega
,A_{\mu }^{h}],\omega ]-\partial _{\mu }\omega +i\frac{g}{2}[\omega
,\partial _{\mu }\omega ]+O(\omega ^{3})\;,  \label{A0}
\end{equation}
One subsequently finds
\begin{equation}
f_{A}[v]=f_{A}[h]+2\mathrm{Tr}\int d^{4}x\,\left( \omega \partial
_{\mu }A_{\mu }^{h}\right) -\mathrm{Tr}\int d^{4}x\,\omega
\partial _{\mu }D_{\mu }(A^{h})\omega +O(\omega ^{3})\;,
\label{func2}
\end{equation}
Armed with this expression, one simply realizes that
\begin{eqnarray}
\delta f_{A}[h] &=&0\;\;\;\Leftrightarrow \;\;\;\partial _{\mu }A_{\mu
}^{h}\;=\;0\;,  \nonumber \\
\delta ^{2}f_{A}[h] &>&0\;\;\;\Leftrightarrow \;\;\;-\partial _{\mu }D{_{\mu }(}%
A^{h}{)}\;>\;0  \label{func3}
\end{eqnarray}
are the conditions for a local minimum. Clearly, this is the a priori reason why the Gribov region $\Omega$, eq.~\eqref{om}, is introduced as it is.

\noindent The transversality condition,
$\partial
_{\mu }A_{\mu }^{h}=0$, can be solved for $h=h(A)$ as a power series in $%
A_{\mu }$. Setting
\begin{equation}
A_{\mu }^{h}=h^{\dagger }A_{\mu }h+\frac{i}{g}h^{\dagger }\partial _{\mu
}h\;,  \qquad
h=e^{ig\xi }=e^{ig\xi^{a}T^{a}}\;,  \label{h0}
\end{equation}
we expand the gauge transformation matrix $h$ in powers of $\xi $
\begin{equation}
h=1+ig\xi -\frac{g^{2}}{2}\xi ^{2}+O(\xi^{3})\;.  \label{hh1}
\end{equation}
As such,
\begin{equation}
A_{\mu }^{h}=A_{\mu }-\partial _{\mu }\xi+ig[A_{\mu },\xi ]+i\frac{g}{2}[\xi ,\partial _{\mu }\xi]+g^{2}\xi A_{\mu }\xi -\frac{g^{2}%
}{2}A_{\mu }\xi ^{2}-\frac{g^{2}}{2}\xi ^{2}A_{\mu }
+O(\xi ^{3})\;.  \label{A1}
\end{equation}
Imposing $\partial _{\mu }A_{\mu }^{h}=0$ yields
\begin{eqnarray}
\partial ^{2}\xi &=&\partial _{\mu }A+ig[\partial _{\mu }A_{\mu },\xi
]+ig[A_{\mu },\partial _{\mu }\xi ]+g^{2}\partial _{\mu }\xi A_{\mu }\xi
+g^{2}\xi \partial _{\mu }A_{\mu }\xi +g^{2}\xi A_{\mu }\partial _{\mu
}\xi   \nonumber \\
&-&\frac{g^{2}}{2}\partial _{\mu }A_{\mu }\xi ^{2}-\frac{g^{2}}{2}A_{\mu
}\partial _{\mu }\xi \xi -\frac{g^{2}}{2}A_{\mu }\xi \partial _{\mu }\xi
-\frac{g^{2}}{2}\partial _{\mu }\xi \xi A_{\mu }-\frac{g^{2}}{2}\xi
\partial _{\mu }\xi A_{\mu }-\frac{g^{2}}{2}\xi ^{2}\partial _{\mu }A_{\mu
}  \nonumber \\
&+&i\frac{g}{2}[\xi ,\partial ^{2}\xi ]+O(\xi ^{3})\;.  \label{hh2}
\end{eqnarray}
Solving iteratively, we arrive at
\begin{equation}
\xi =\frac{1}{\partial ^{2}}\partial _{\mu }A_{\mu }+i\frac{g}{\partial ^{2}%
}\left[ \partial A,\frac{\partial A}{\partial ^{2}}\right] +i\frac{g}{%
\partial ^{2}}\left[ A_{\mu },\partial _{\mu }\frac{\partial A}{\partial ^{2}%
}\right] +\frac{i}{2}\frac{g}{\partial ^{2}}\left[ \frac{\partial A}{%
\partial ^{2}},\partial A\right] +O(A^{3})\;,  \label{phi0}
\end{equation}
and thus
\begin{eqnarray}
A_{\mu }^{h} &=&A_{\mu }-\frac{1}{\partial ^{2}}\partial _{\mu }\partial A-ig%
\frac{\partial _{\mu }}{\partial ^{2}}\left[ A_{\nu },\partial _{\nu }\frac{%
\partial A}{\partial ^{2}}\right] -i\frac{g}{2}\frac{\partial _{\mu }}{%
\partial ^{2}}\left[ \partial A,\frac{1}{\partial ^{2}}\partial A\right]
+ig\left[ A_{\mu },\frac{1}{\partial ^{2}}\partial A\right] +i\frac{g}{2}%
\left[ \frac{1}{\partial ^{2}}\partial A,\frac{\partial _{\mu }}{\partial
^{2}}\partial A\right] +O(A^{3})\;.  \label{minn2}
\end{eqnarray}
It is interesting to rewrite $A_\mu^h$ as
\begin{eqnarray}
A_{\mu }^{h} &=&\left( \delta _{\mu \nu }-\frac{\partial _{\mu }\partial
_{\nu }}{\partial ^{2}}\right) \left( A_{\nu }-ig\left[ \frac{1}{\partial
^{2}}\partial A,A_{\nu }\right] +\frac{ig}{2}\left[ \frac{1}{\partial ^{2}}%
\partial A,\partial _{\nu }\frac{1}{\partial ^{2}}\partial A\right] \right)
+O(A^{3})  \nonumber \\
&=&\left( \delta _{\mu \nu }-\frac{\partial _{\mu }\partial
_{\nu }}{\partial ^{2}}\right) \Psi_\nu  \label{hhh3}
\end{eqnarray}
Under an infinitesimal gauge transformation
\begin{equation}
\delta A_{\mu }=-\partial _{\mu }\lambda +ig[A_{\mu },\lambda ]\;.
\label{gauge3}
\end{equation}
it can be checked that
\begin{equation}
\delta \Psi _{\nu }=-\partial _{\nu }\left( \lambda -i\frac{g}{2}\left[ \frac{%
\partial A}{\partial ^{2}},\lambda \right] \right) +O(\lambda^2)\;,  \label{phi1}
\end{equation}
The combined knowledge of \eqref{hhh3} and \eqref{phi1} nicely displays that $A_\mu^h$ is indeed transverse, while it is also gauge invariant, order by order. It is perhaps interesting to notice here that in \cite{Gracey:2007ki}, the one loop renormalizability of the non-local operator $\frac{1}{2}\int d^4x A^ h A^h$, i.e.~the local minimum of eq.~\eqref{norm1}, was explicitly checked.
\end{widetext}

\end{document}